# A Cache Management Strategy to Replace Wear Leveling Techniques for Embedded Flash Memory

Jalil Boukhobza[*], Pierre Olivier[*], Stéphane Rubini[+]
Université Européenne de Bretagne, France
Université de Brest ; CNRS, [*]UMR 3192 Lab-STICC, [+]EA3883 LISyC
20 avenue Le Gorgeu
29285 Brest cedex 3, France
{boukhobza, rubini}@univ-brest.fr, pierre.olivier@etudiant.univ-brest.fr

*Abstract*— **Prices of NAND flash memories are falling drastically due to market growth and fabrication process mastering while research efforts from a technological point of view in terms of endurance and density are very active. NAND flash memories are becoming the most important storage media in mobile computing and tend to be less confined to this area. The major constraint of such a technology is the limited number of possible erase operations per block which tend to quickly provoke memory wear out. To cope with this issue, state-of-the-art solutions implement wear leveling policies to level the wear out of the memory and so increase its lifetime. These policies are integrated into the Flash Translation Layer (FTL) and greatly contribute in decreasing the write performance. In this paper, we propose to reduce the flash memory wear out problem and improve its performance by absorbing the erase operations throughout a dual cache system replacing FTL wear leveling and garbage collection services. We justify this idea by proposing a first performance evaluation of an exclusively cache based system for embedded flash memories. Unlike wear leveling schemes, the proposed cache solution reduces the total number of erase operations reported on the media by absorbing them in the cache for workloads expressing a minimal global sequential rate.**

*Keywords-component; NAND Flash memory; cache; FTL; performance; simulation; storage; I/O workload; wear levelling*

I. INTRODUCTION

NAND flash memories are more and more used as main storage systems. We can find them in music players, smart phones, digital cameras, laptop computers, and a huge set of electronic appliances. NAND flash memory is based on semiconductor chips giving them some extremely interesting characteristics. They are small, lightweight, shock resistant, and very power efficient. One of its main constraints is its poor lifetime which is due to the limited number of erase operations one can perform on a given block (see background section). In addition to this limitation, flash memories present poor write performance because writing can be performed only in one direction (from 1 to 0), so a block must be erased before being modified.

The FTL (Flash Translation Layer) is a hardware/software layer implemented inside the flash-based device. One of its main functionalities is the mapping of logical addresses coming from higher levels (applications/operating system) to lower level physical addresses. Throughout this mapping is achieved the wear leveling, which is the widely adopted technique to cope with the limited number of erase operations. Wear leveling consists in spreading the write operations over the whole flash memory area to increase the block average lifetime preventing it from wearing out too quickly. So, in order to modify a given block, it is rewritten to another free one that suffered less erase cycles. The use of such a technique implies to take into account the validity of written data as many versions can be present in different blocks. Garbage collectors are implemented to recover free space when needed.

Wear leveling in FTL relies on an on-flash SRAM-based memory where the logical-to-physical mapping tables are stored. The use of such an SRAM being very expensive for embedded systems, FTL designers tried to severely minimize the size of such metadata. In fact, one have to find a trade-off between increasing the performance of the flash memory by allowing a page mapping algorithm, which consumes a lot of SRAM but reduces the number of erase operations, and reducing the SRAM usage by reducing the metadata and so increasing the granularity of the mapping which increases the number of erase operations. All state of the art solutions are located between those two extremes.

In addition to minimizing the standard deviation of the erasure distribution over the flash media, a good wear leveler should minimize extra erase operations due its own execution. Despite this, a wear leveler cannot lower the total number of erase operations under a certain limit. That is why researchers designed buffering systems on top of FTLs [1-4]. They are mainly used to reorganize non sequential request streams which contributes in minimizing the number of erase operations in addition to buffering the data for future use. The main disadvantages of those systems are: 1) the price: too high for many embedded systems like mp3 players, flash memory cards, etc., 2) the complexity: two levels of complexity are cumulated, FTL and cache, and 3) a higher CPU usage.

As stated in [5], the endurance problem of flash memories has been widely overestimated and actual flash memories, thanks to recovery time, can sustain much more erase operations than expected.

Because of all the reasons stated above, we think that FTL solutions are temporary and that one should begin considering exclusively cache based flash memories mainly for embedded systems.

We propose, in this paper, a solution based on a simple flexible dual cache to replace wear levelers and garbage collectors. This exclusively based cache system is compared to state-of-the-art FTLs to show that we can rely upon a cache system to minimize and wear level the erase operations and improve performances. This cache system performs well for most tested workloads except for extremely random workloads with very small request sizes.

The main research contributions of this study are the following: 1) a novel caching technique to replace the existing wear leveling (and garbage collection) solutions. 2) The implementation of a simulator used to achieve performance evaluation based on FlashSim [6][7] that is built by extending DiskSim 3.0 [8], the most popular storage system simulator. We added a cache level support for FlashSim. 3) The validation with a large number of synthetic workloads based on global sequentiality parameter.

The paper is organized as follows: a description of flash memories is given in the background section and some related works are discussed in the third section. In the forth section, the proposed cache system organization and policies are depicted. The following section gives performance evaluation methodology and discusses the simulation results, and finally we conclude and give some perspectives in the last section.

## II. BACKGROUND

Flash memories are nonvolatile EEPROMs. They are of mainly two types: 1) NOR flash and 2) NAND flash, they are named after the logic gates used as the basic structure for their fabrication. NOR flash memories support bytes random access and have a lower density and a higher cost as compared to NAND flash memories. NOR memories are more suitable for storing code [9]. NAND flash memories are, by contrast, block addressed, but offer a higher bit density and a lower cost and provide good performance for large read/write operations (as compared to NOR memories). Those properties make them more suitable for storing data [9]. The study presented in this paper concerns NAND flash memories.

NAND flash memories can be classified into two categories: 1) Single Level Cell (SLC) and 2) Multi Level Cell (MLC). In SLC flash memories, only one bit can be stored in one cell, while in MLC, two bits or more can be stored. The price to pay when acquiring MLC as compared to SLC technology is quite high. SLC can have up to 10 times higher lifetime and lower access latency [10].

Flash memory is structured as follows: it is composed of one or more chips; each chip is divided into multiple planes. A plane is composed of a fixed number of blocks, each of them encloses a fixed number of pages that is multiple of 32. Actual versions of flash memory have between 128 and 1024 KB blocks (pages of 2-8KB). A page actually consists of user data space and a small metadata area [7][2].

Three key operations are possible on flash memories: read, write and erase. Read and write operations are performed on pages, whilst erase operations are performed on blocks. Pages in a given block must be written sequentially. NAND flash does not support in-place data modification. In order to modify a page, it must be erased and rewritten in the same location or completely copied to another page and its corresponding LBA-PBA translation map entry is updated.

One fundamental constraint on flash memories is the limited number of write/erase cycles (from $10^4$ for MLC to $10^5$ for SLC) [10]. After the maximum number of erase cycles is reached, a given memory cell can no more retain data. Some spare storage cells exist on the chip to cope with such a wearing out. Due to data locality (temporal and/or spatial in many workloads), some of those memory blocks would be much more used than the others. They then tend to wear out more quickly. This very central problem pushed the researchers' community to seriously consider wear leveling techniques to even out the write operations over the whole memory blocks even though these techniques reduce dramatically the performance.

As mentioned earlier, writing a data requires having a clean page available (free page from a block erased beforehand). If the system has to modify a previously written page, it looks for a clean page and writes the data while invalidating the previously occupied page (by modifying the metadata). The erase operation is done asynchronously. If there are no clean pages, a garbage collector is ran to recycle invalidated pages.

## III. RELATED WORK

### A. FTL Based Systems

One can distinguish three main FTL schemes, named after the granularity of the logical to physical mapping. 1) Page mapping provides great flexibility, at the cost of a large mapping table. 2) Block mapping increases the mapping granularity to a block level, reducing drastically the size of the mapping table. When updating data in a previously written page, the system must copy the entire block to a new free location, leading to huge performance drops. 3) Hybrid mapping tries to mix both page and block mapping techniques. Flash space is partitioned into data blocks, mapped by blocks and log blocks, mapped by pages. When the number of free log blocks becomes low, the system must merge them to create data blocks. Merge operations are extremely costly. Current FTL systems are derived from those three main schemes.

Demand-based Flash Translation Layer (DFTL) [7] is a page mapping-based FTL. Only a part of the mapping table is present in SRAM, most of it residing on the flash media itself. Based on DFTL, Convertible Flash Translation Layer (CFTL) [11] also places the major part of its page mapping table on the flash memory. CFTL separates data into cold and hot data, respectively mapped by block and by pages.

Regarding hybrid-based schemes, Dynamic dAta Clustering (DAC) [12] separates the flash space into multiple regions, from the coldest to the hottest data. This helps to avoid situations with blocks having mixed hot and cold data which lead to costly full merges. Fully Associative Sector Translation (FAST) [13] tries to increase the number of non-costly switch merges by partitioning the log blocks into sequentially and randomly written log blocks, sequentially written blocks being good candidates for switch merge operations.

Block mapping-based scheme Mitsubishi [14] [15] and M-Systems FMAX algorithm [14] divide flash space into 1) normally mapped blocks in which physical pages offset is always equal to the corresponding logical page offset, and 2) additional blocks written independently of the offset. In order to retrieve data written in additional blocks, the corresponding logical page number is written in the Out-Of-Band area of each physical page.

Actual FTL algorithms try to bypass the limitations of the main scheme they derivate from, by storing the mapping table on the flash media for page mapping based schemes, reducing the number of costly full merge operations for hybrid-based schemes, and delaying the block copy operations on page update for block mapping-based algorithms.

For the sake of this study, we have chosen to compare our system to two efficient FTL schemes that are DFTL (page mapped) and FAST (hybrid).

### B. Buffers above FTL Systems

Even though designed FTL techniques are more and more efficient (and complex), performance of write operations are still very poor. Buffering systems have been designed to cope with this issue by reorganizing non-sequential request streams before sending them to the FTL. Those buffers are placed on top of FTLs.

The idea behind Flash Aware Buffer (FAB) [1] is to keep a given amount of pages in the buffer and flush pages that belong to fullest block. The Cold and Largest Cluster policy (CLC) [2] system implements two clusters of pages in the cache, a size-independent and a size-dependant one based on LRU. Block Padding Least Recently Used (BPLRU) is a write buffer [3] that uses three main techniques: block level LRU, page padding and LRU compensation. Block-Page Adaptive Cache (BPAC) [4] partitions the cache into two parts: an LRU page list used to store data with high temporal locality, and a list of blocks divided itself into two data clusters organized by recency and size.

Most of those buffer techniques have been designed to be implemented on SSD caches which implies two major characteristics: 1) they rely on too big size buffers (8 to 128MB) and so are not adequate for some embedded systems, 2) they are implemented on top of FTLs for which they do not know the details. This is an advantage if one wants to use the buffer on an existing system (which is the purpose of the authors), but can be a limitation if one wants to design a well performing system.

## IV. C-LASH ARCHITECTURE FOR EMBEDDED STORAGE

In this section we describe C-lash (Cache for fLASH) system architecture and policies. C-lash is much smaller than previously discussed cache systems (512KB for the evaluations listed in this paper).

### A. C-lash Architecture

In C-lash, the cache space is partitioned into two spaces, a page space (p-space) and a block space (b-space). P-space consists of a set of pages that may come from different blocks in the flash while b-space is composed of blocks that can be directly mapped (see Fig. 1).

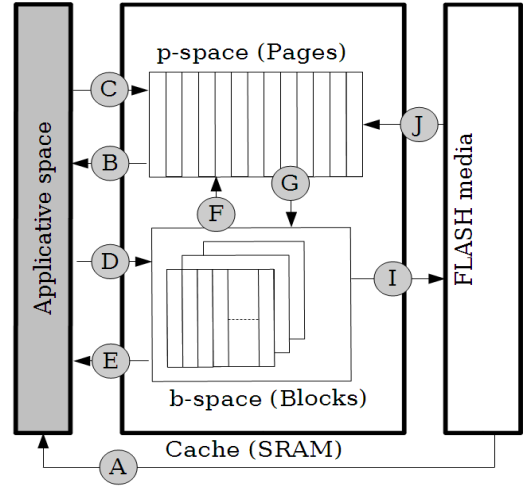

Figure 1. Structure of the C-lash system.

C-lash is also hierarchical, it has two levels of eviction policies: one which evicts pages from p-space to b-space (G in Fig. 1), and another level in which blocks from b-space are evicted into the flash media (I in Fig. 1). With this scheme, we insure that the system always flushes blocks rather than pages to the flash memory. P-space and b-space regions always contain respectively either valid or free pages or blocks.

When a read request arrives, data are searched in both spaces. If we get a cache hit, data are read from the cache (B or E in Fig. 1), otherwise, they are read from the flash memory (A in Fig. 1). A read miss does not generate a copy into the cache.

When a write operation is issued, in case of a cache hit, data are overwritten (respectively C or D in Fig. 1) with no impact on the flash media. If data are not in the cache, they can only be written into the p-space (C in Fig. 1). If enough pages are available, we use them to write the data. If not, we choose some pages to flush from the p-space to the b-space (G in Fig. 1) and copy the new data.

### B. Cache Policies Algorithms

Two eviction policies are implemented for each of p-space and b-space (respectively G and I in Fig. 1).

#### 1) P-space Eviction Policy

When a write request is issued and the considered page is neither present in the p-space nor in the b-space, a new page must be allocated in the cache p-space. If a free page is available, the new page is written and data in the corresponding location in the flash invalidated. If no free space is available, the system chooses a set of pages to evict into the b-space (not into the flash media). Pages to evict are those forming the largest set from the same block. Once this set found, we have two possibilities:

1. A free block is available in the b-space and so victim pages are copied in it.

2. No free block is available and then, the set of victim pages is compared to the number of valid pages contained in each block of the b-space area:

   a. If there exists a block containing less valid pages than the victim ones, a switch operation is performed (F and G in Fig. 1): pages in the victim block are moved in the p-space while the subset of victim pages is moved in the freed block. This induces no impact on the flash memory.

   b. If all blocks in the b-space contain more valid pages than the subset of victim pages: the b-space eviction policy is executed to flush a block into the flash media.

In the upper illustration of Fig. 2, we describe an example of a p-space eviction; the chosen pages are those belonging to block B21 containing the biggest number of pages. One block in the b-space contains two valid pages which is less than the 3 pages to evict. Thus, C-lash switches between the block B4 and the three pages subset. The system, therefore, frees one page without flushing data into the flash.

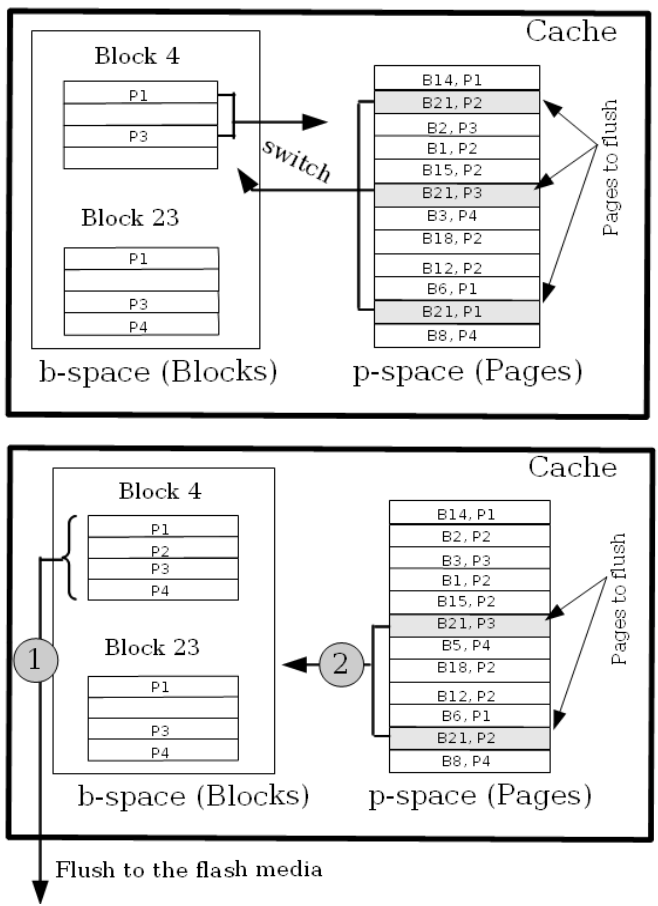

Figure 2. Two examples describing different scenarios of the eviction policy for p-space and b-space.

*2) B-space Eviction Policy*

An LRU algorithm is used for this eviction to take into consideration, in the b-space, the temporal locality exhibited by many workloads. When a block eviction is performed, the whole corresponding block in the flash memory is erased before being replaced by the one in the cache.

In case the flash memory still contains some valid pages of the block to evict from the cache, a merge operation (J in Fig. 1) is to be performed. This merge operation consists in reading the still valid pages in the flash memory before flushing the whole block from the cache. The merge can be done either during a p-space eviction, we call it *early merge*, or just before flushing the block on the flash media, we call it *late merge*. Early merge is more advantageous if the workload is read intensive and shows temporal and/or spatial locality (make benefit from the cached data). If the workload is write-intensive, we would use the late merge. By delaying the merge operation we insure two main optimizations: 1) we read a minimum number of pages since between the moment the pages are evicted into the b-space and the moment they will be evicted into the flash, many pages can be written and so invalidated from the flash. 2) Since it is possible for a block in b-space to be moved to p-space during a p-space pages eviction (a switch), it may not be worth doing the merge operation too early. This can cause extra read operations. We restrict the scope of the presented study to the use of a late merge scheme.

An example of block eviction is shown in Fig. 2. In the lower illustration, we describe a p-space eviction leading to a b-space flush to the flash media. In the p-space eviction phase, two pages of the B21 block are chosen to be evicted. The blocks in the b-space contain more valid pages than the pages amount to be evicted from the p-space. So, the system needs to evict a block into the flash beforehand. After that, the system copies both pages of the B21 into the freed block. In this specific example, no merge operation occurs because the flushed block is full of valid pages.

V. PERFORMANCE EVALUATION

In this part, we compare the C-lash system and three efficient state-of-the-art FTLs: DFTL, FAST and the idealized page mapping algorithm used as a baseline. The purpose of the performance evaluation part is to study if caching techniques can replace actual efficient FTL schemes for embedded storage.

*A. Simulation Framework and Methodology*

FlashSim is a simulator based on Disksim [8] which is the most popular disk drive storage system simulator in both academia and industry. Disksim is an event driven simulator that has been validated with an exhaustive set of disk models. It simulates the whole storage system going from the device driver until the detailed disk movement, integrating many controller strategies, queues, caches, bus systems, and detailed disk descriptions. Disksim also integrates a very detailed synthetic workload generator which is used in part of our experimentations. Disksim does not natively include flash memory support.

FlashSim integrates some modules that are specific to the flash memory subsystem simulation. In addition to the components modeled by Disksim, it is able to simulate the basic flash device infrastructure to implement specific operations: read, write, erase, etc. Logical to physical address translation mechanisms are also implemented with garbage collection policies. FlashSim implements many FTL schemes: the FAST, DFTL schemes and an idealized page based FTL.

We have modified and increased the functionality of FlashSim to allow the simulation of a dual cache subsystem placed on the top of the flash media. The used cache is configurable as many cache policies can be simulated in the b-space (FIFO, LRU, and LFU). Only the study based on LRU is shown in this paper.

### B. Storage System and Performance Metrics

We rely on three main performance metrics: the average request response time, the number of performed erase operations and weighted standard deviation of the erase operations distribution to express the quality of wear leveling. In our study, the response time is captured from the I/O driver point of view, including all intermediate delays: caches, controllers, I/O queues, etc. We tried to minimize the intermediate elements' impact to focus on the flash memory subsystem behavior. The second metric we capture is the number of performed erase operations. It indicates the wear out of the memory.

We simulated a NAND flash memory with a 2KB page size and a 128KB block size. The three operations have the following delays: 130.9μs for a page read: 405.9μs for a page write, and 2ms for a block erase: [16]. The chosen C-lash configuration has 2 blocks (128KB each) in the b-space and 128 pages in the p-space (a total size of 512KB). The mapping table used with C-lash is extremely small (much less than 1KB) as compared to the cache size. All the performed simulations begun with the same initial state, the flash media was supposed to be completely dirty. Each new write in a given block generated an erase operation.

### C. Simulated Workloads

The performed tests are those described in Table 1. We mainly vary the global sequentiality of the workload by tuning three parameters: sequentiality (strictly contiguous requests), spatial locality (interleaved contiguous requests [8]) and request size.

We also varied the number of requested data which gives an idea on the size of used address space and so the free area of the flash. This last parameter is important because for some FTL schemes, the more free space is available, the better are the performances. Nevertheless, for embedded systems, it is more realistic to consider that the flash memory space has been optimized and then, there is not much free space available.

**Table 1.** Tested workloads characteristics on 1GB I/O space (default values are underlined unless otherwise specified).

| Seq. rate | Spatial locality | Mean Req. Size |
|---|---|---|
| **0**→100% (steps of 5%) | **0**→100% (steps of 10%) | 1, 2, **4**, 8, 16, 32, 64, 128 pages |
| **Request number** | **Write rate** | **Inter arrival times** |
| 60000 | 80% | exp (0, 200ms) |

### D. Results & Discussion

In this section, we describe the results obtained when comparing our cache for flash solution with the following FTLs: FAST, DFTL and an idealized page mapping FTL. As stated in the background section, the pure page mapping FTL is very RAM consuming and is not really usable, but it gives the best FTL performance. As in [7], we use it as a baseline in our study.

*1) Sequentiality:* As we can see in Fig.3, C-lash performs better than DFTL, in terms of response time, if the sequentiality rate is above 40 %. It always performs better than FAST.

It achieves less erase operations if the sequentiality rate is above 50% and always outperforms FAST.

Thanks to the dual cache eviction properties of C-lash which help absorbing the erase operations, C-lash always perform a better wear leveling than both FAST and DFTL.

The more the sequentiality rate is high, the more C-lash approaches the ideal page mapping FTL performance for the three performance metrics.

*2) Spatial Locality.* We can draw the same conclusions concerning the spatial locality as for the sequentiality. In terms of response time, C-lash always acheives better performance than FAST, it performs better than DFTL if spatial locality is above 10%, and performs even better than the page mapping FTL if spatial locality is greater than 85%.

Concerning the number of erasures, C-lash takes advantage of its buffering mechanism to reveal sequentiality and so performs better than DFTL for more than 20% spatial locality, and better than page map FTL for more than 85% spatial locality.

Spatial locality defines requested data that are neighbors but not strictly contiguous. Two requests, R1 and R2, are supposed to be neighbors if the distance between the start address of R1 and the start address of R2 is under a given threshold. We defined, for the sake of these tests, the threshold to be less than two pages (we used a normal distribution). That is the reason why for 100% spatial locality C-lash gives far better performance than other FTLs, it is because C-lash takes benefit of both spatial and temporal locality (through the Disksim spatial locality parameter, we can also simulate temporal locality if the distance is nil).

We can observe that the weighted standard deviation increases for 100% spatial locality. For the three FTLs, this is mainly due to the very high temporal locality making the same data blocks to be very frequently modified. For the case of C-lash, the number of erase operations is too small (write/modify absorbed by the cache) to consider the case of 100% spatial locality. In fact, we have less than 10 erase operations. This case is not representative of real workloads.

*3) Request Size:* These tests were performed with 0% sequentiality and 20% spatial locality. We see that for requests sizes greater than 6 pages (of 2KB), C-lash mean response time is better than DFTL mean response time. C-lash even

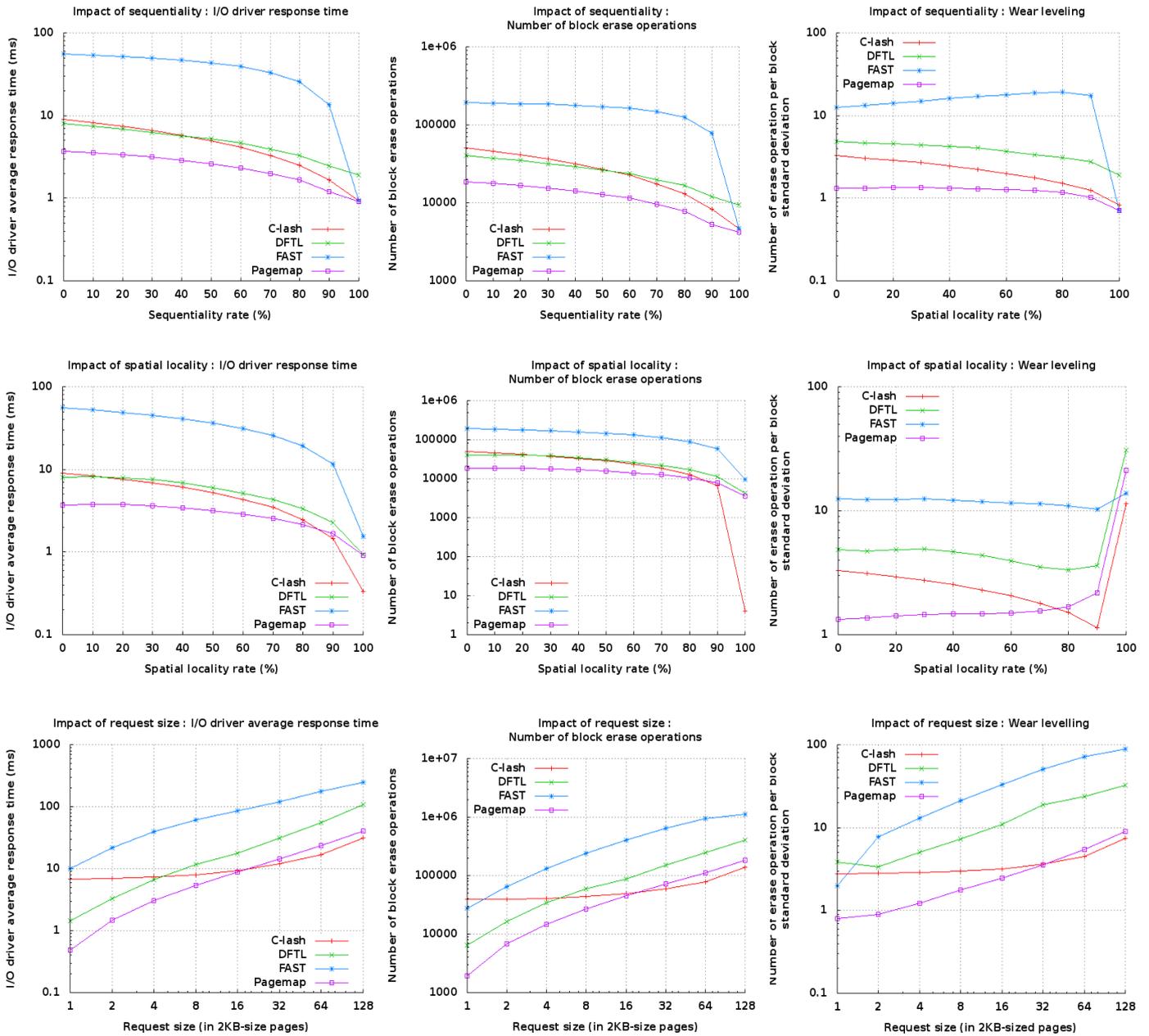

Figure 3. Performance evaluation for C-lash according to sequential rate, spatial locality and request size.

performs better than the page mapping FTL for request sizes greater than 16 pages.

We can observe the same behavior for the number of erasures.

We can also notice that the slope of both response times and number of erasures curves for the three FTLs is steeper (logarithmic scale in the figure) than the one of C-lash. In fact, C-lash takes more benefit from the increasing request sizes. The more the request size is important, the less we have to do costly merge operations.

In fact, the request size is just another facet of sequentiality. The bigger the request size is, the more we access to contiguous pages into a given block of the flash memory, and so better is the throughput.

*4) Request number:* For this study, we fixed the request size to 6 pages. As we can see in Fig 4., we have varied the number of requests in two modes: a random workload and a sequential one.

Generally, we can observe that C-lash performance is independent from the number of issued requests. In fact, this parameter has no impact on its performance in terms of response time/throughput.

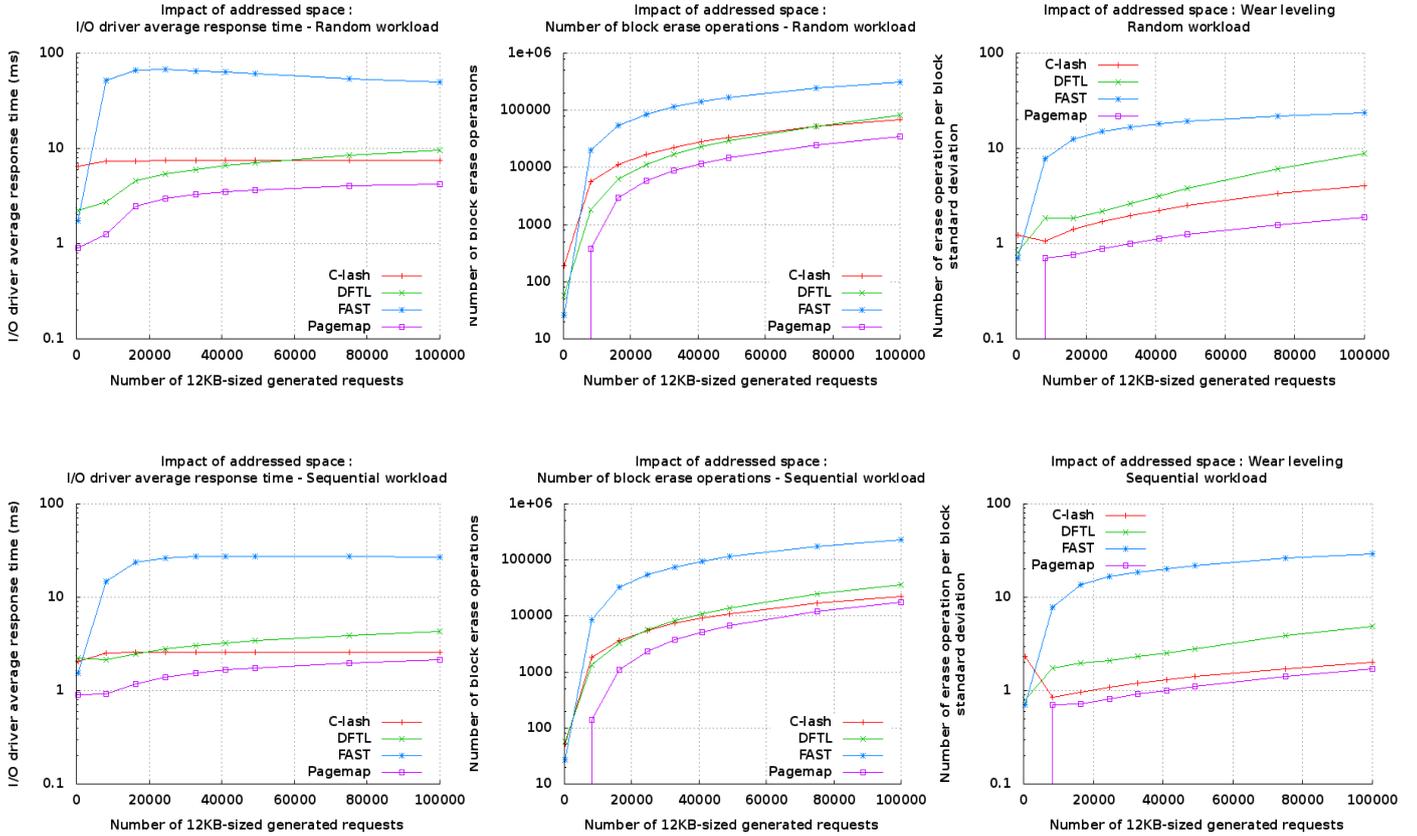

Figure 4. Performance evaluation for C-lash according to the number of issued I/O requests.

FTLs take more benefit from the available free space in the flash media to delay erase operations. For DFTL and page mapping FTLs, the more requests we have, the worse is the performance. For random workloads, C-lash outperforms DFTL when more than 60000 requests are issued, which approximately represents 700MB of data (from 1GB space). For sequential accesses, C-lash outperforms DFTL earlier: when there is more than 20000 requests issued, which represents 230MB of data. The same conclusions can be drawn for the number of erasures: the more I/O requests we issue, the better is the performance of C-lash as compared to other FTLs. Wear leveling performed by C-lash is always better than DFTL and FAST for the tested workloads except for extremely small number of requests, which is not realistic.

In fact, the number of requests gives us an idea of the total congestion of the flash memory. Performance of C-lash is independent from this parameter while the performances of the different FTLs highly depend on the available free space on the flash media to perform wear leveling. Most embedded systems do not oversize their secondary storage for cost reasons. That is why we think that one does not have to count on the amount of free space available to better absorb erasures.

## VI. LIMITATIONS OF C-LASH

• C-lash system performs poorly for very random workloads with small request sizes but still performance drop is not dramatic and we can notice that many workloads are sequential for an important number of embedded systems (mp3 and mp4 players, cameras, etc).

• In case of a power failure, data in the cache can be lost. This is a very important problem especially for embedded systems; a possible solution consists in adding a small battery [3] allowing to flush all the valid pages of the cache in the flash media. In C-lash system, we thought about this problem and we implemented the LRU algorithm on the b-space and rather than on the p-space which would have provoked more flush operations on the media. This ensures a more regular data flush operations, and so less data loss, sometimes at the expense of performance (increase of response time).

• For some FTLs like DFTL, the more free space is available, the better are performances. C-lash system performance can lag behind even when sequential rate is somewhat high if flash memory is supposed to be nearly free. This case did not seem realistic to us, that is why performed tests does not consider this case even though the request number could have been increased for sequential rate tests.

## VII. CONCLUSION

We think that the FTL must be considered as a temporary solution, as flash-based embedded storage systems could migrate in a near future toward exclusively cache-based solution. This migration will unload the flash subsystem from more and more complex wear leveling and garbage collection schemes, and so reducing FTL overhead and improving the overall storage system performance.

In this paper, we presented C-lash, an exclusively cache-based solution that outperforms state-of-the-art FTL schemes performance under workloads showing a minimum global sequentiality. As stated in this paper, sequentiality is characterized by three parameters: the sequentiality rate for strictly contiguous requests, the spatial locality for neighbor /local requests and the request size.

The performance evaluation of this study also showed that C-lash performs a better wear leveling than the tested FTL solutions (DFTL and FAST) by showing a less important weighted standard deviation. Good wear leveling performances is achieved thanks to the LRU scheme implemented at the b-space level which insures the cache to absorb erasures caused by temporal locality exhibited by many workloads.

While C-lash architecture shows good performances with embedded constraints (size, congestion, complexity) with a large set of workloads, it is not completely adequate for large scale storage systems like SSDs. As an extension perspective, we are studying a composite structure including a FTL and a collaborative cache based on C-lash. The FTL, in this case, is only used to deal with random workloads. Indeed, while FTL systems can be replaced by caching mechanisms for embedded systems, we think that they are still useful to manage random workloads in mass storage systems in collaboration of well performing caching mechanism.

The Disksim simulator including the C-lash support will soon be available online with the bunch of performed tests.